# DiscopFlow: A new Tool for Discovering Organizational Structures and Interaction Protocols in WorkFlow

Mahdi ABDELKAFI, Lotfi BOUZGUENDA, Faiez GARGOURI
Mir@cl laboratory/ University of Sfax/ISIM
Route de Tunis, km 10.
BP 242, 3021 Sakeit Ezzit, Sfax-Tunisia.
Emails: Abdelkafi@hotmail.fr, {lotfi.bouzguenda, faiez.gargouri}@isimsf.rnu.tn

*Abstract*—This work deals with Workflow Mining (WM) a very active and promising research area. First, in this paper we give a critical and comparative study of three representative WM systems of this area: the ProM, InWolve and WorkflowMiner systems. The comparison is made according to quality criteria that we have defined such as the capacity to filter and convert a Workflow log, the capacity to discover workflow perspectives and the capacity to support Multi-Analysis of processes. The major drawback of these systems is the non possibility to deal with organizational perspective discovering issue. We mean by organizational perspective, the organizational structures (federation, coalition, market or hierarchy) and interaction protocols (contract net, auction or vote). This paper defends the idea that organizational dimension in Multi-Agent System is an appropriate approach to support the discovering of this organizational perspective. Second, the paper proposes a Workflow log meta-model which extends the classical one by considering the interactions among actors thanks to the FIPA-ACL Performatives. Third, it describes in details our DiscopFlow tool which validates our contribution.

## I. INTRODUCTION

The Workflow is a key technology which aims at automating the coordination of activities composing business process [1]. A Workflow Management System (WfMS for short) is a software which permits to define, implement and perform one or several business processes. The traditional interest of a WfMS focuses on design, configuration and enactment phases [2]. As a consequence, there are a few supports for diagnosis phase. Besides, the support for the design phase is limited to provide a simple editor yet that an analysis is useful for the design which is neglected. A very few WfMS propose the simulation, the checking of design step and also support to help the interpretation of data from execution traces. Although the majority of actual WfMS collect the business instances execution traces, they do not propose a support to exploit this important information. In this context, the Workflow Mining (WM) area has recently appeared and considered as a current research area. More precisely, the WM aims mainly at analyzing the workflow execution traces (or Workflow log) in order to discover the Workflow perspectives such as the Organizational Perspective (OP), the Informational Perspective (IP) and the Process Perspective (PP), which help the monitor to improve or propose a new workflow [3]. The WM is considered as an appropriate method to support Business Processes Reverse engineering (BPR).

Most of the proposed workflow mining systems such as ProM [4], InWolvE [5] and WorkflowMiner [6] only focus on the process perspective discovering. The organizational perspective discovering is not considered in the existing WM systems except the ProM system. This latter supports only the discovering of actors, their roles and five kind of social network (Handover of work, Subcontracting, Working together, Reassignments, Doing similar tasks) [2]. We show later that the discovering of organizational perspective is not limited to the previous elements but also it includes others kind of social networks/Organizational Structures (OS) such as federation, coalition, market or hierarchy and Interaction Protocols (IP) such as contract net, vote or auction as well defended in Multi-Agent System (MAS) area [7].

To the best of our knowledge, the reason why the existing propositions only deal with process perspective mining because they use a workflow log limited to the activities execution, actors who execute these activities and no traces on the interactions among actors.

The problem being addressed in this paper is: "*how to discover the Organizational Structures (OS) and Interaction Protocols (IP) from workflow log which also integrates exchange/interaction between actors?*"

In this paper, our main goal is in the one hand to give a comparative study of WM systems according to the defined quality criteria in order to emphasis the insufficient of organizational perspective discovering. On the second hand, we show how MAS approach can deal with the major drawback of the studied systems and namely the discovering of the OS and IP as mentioned previously.

The remainder of this paper is organized as follows. Section 2 (i) presents the quality criteria that we have defined and (ii)

compare some existing WM systems according to these quality criteria. Section 3 describes the proposed Workflow log meta-model which extends the classical one. It starts by justifying the interest of organizational dimension in MAS approach to deal with organizational structures and interaction protocols discovering. Then, it presents our proposed workflow log meta-model. Section 4 presents in detail our DiscopFlow tool which validates our solution. Section 5 concludes the paper and outlines the main perspectives of this work.

## II. A COMPARATIVE STUDY OF SOME EXISTING WORKFLOW MINING SYSTEMS

The purpose of this section is to define the criteria to measure the quality of a WM system and compare some existing systems, representative of the state of the art, according to these quality criteria. More precisely we compare the ProM [4], InWolve [5] and WorkflowMiner [6] systems. Note that the existing systems such as EmiT, Thumb, MinSocN and MiMo are merged within the ProM system. For more information on comparative study of existing WM systems we refer to [8].

### A. Quality criteria

The most important quality criteria are the following:

- *the capacity to filter and convert a Workflow log,*
- *the capacity to discover workflow perspectives,*
- *the capacity to support Multi-Analysis of processes.*

Let us detail each quality criteria.

**The capacity to filter and convert a Workflow log.** The filtering concerns mainly the separation between the principal activities and the optional activities known logistics activities and the taking into account only the completed activities. Often, the noise being in workflow log exists at ad-hoc workflow systems and groupware product which are based on unstructured process activities. The conversion concerns the transformation of a workflow log from a given format to other (for instance from text to XML) to ease the extraction process of information and then to analyze it.

**The capacity to discover workflow perspectives.** Regarding the informational perspective, the ability to discover the consumed documents and produced documents by a workflow. Regarding the organizational perspective, the ability to discover clearly the actors, their roles, their organizational units, the policy of activities allocation (i.e. the employed interaction protocols between actors and social networks describing the nature of collaborations between actors (federation, coalition, hierarchy, market and so on). Regarding the process perspective, the ability to discover clearly the activities, workflows patterns (sequential, parallel, iterative and so on) and their coordination.

**The capacity to support Multi-Analysis of processes.** The possibility given to designers to make various analysis such as **Delta Analysis** and **Performance Analysis.** The analysis of performances concerns mainly the analysis of performances of a workflow component such as activity, actors,… then to proceed to the modification of existing models. According to the literature [9], four performance metrics have been proposed for process perspective such as flow time, waiting time, processing time and synchronization time. Regarding organizational perspective, four metric have been also proposed such as frequencies, time, utilization and variability. The Delta analysis consists of comparing the prescribed processes (i.e. theory) and the deployed processes (i.e. practice). This analysis of differences permits to proceed then to adjust and/or enhance the processes. It also allows the comparison of different implementations of process within various organizations.

Besides to these criteria, other ones must be taken into account and which are related to the quality of software in general such as **Usability of interfaces**, **Portability** and **Extensibility**. Regarding the usability of interfaces, the ability of the system to allow designers (i) to model clearly the workflow models with graphic representation (Petri Nets for instance) and (ii) to make simulations and animations for detecting some errors and ambiguities.

### B. Comparative study of some existing WM systems

Because of space constraint, we give here only a comparative table of studied WM systems.

This comparative study calls three remarks:

- The capacity to filter and convert a workflow log is assured by all the WM systems expect the WorkflowMiner which does not support the filtering of workflow log.

- The capacity to discover the three workflow perspectives is not supported in totality by the WM systems. The InWolvE and WorkflowMiner systems support only the discovering of process perspective while the ProM system supports the process perspective and organizational perspective in terms of actors and their roles (see introduction). The informational perspective is neglected by all the systems.

- The capacity to support multi-analysis is ensured only by the ProM system while the InWolve system does not support any analysis and the WorkflowMiner supports only the analysis of performance.

As mentioned in introduction, the reason why the existing propositions only deal with process perspective mining because they use a workflow log limited to the activities execution, actors whose execute these activities and no traces on the interactions among actors. However, we have defined our own Workflow log meta-model which allow the discovering of the three complementary workflow perspectives and more particular the discovering of organizational structures and interactions protocols. We have also developed our tool which fulfils the previous quality criteria.

TABLE I. A COMPARATIVE STUDY OF WM SYSTEMS.

| Existing WM systems / Quality criteria | | | ProM [van Dongen, 05] | InWolve [Herbst, 03] | WorkflowMiner [Baïna, 06] |
|---|---|---|---|---|---|
| Pre-Handling of Workflow log | | Filtering | Log Filter | + | - |
| | | Conversion | From EPC to Petri Nets | From XML to APF or ASCII | Adapters |
| Capacity to discover Workflow Perspectives | | Informational Perspective | - | - | - |
| | Organizational Perspective | Actors | Plug-in originator by task matrix) | - | - |
| | | Roles | + | - | - |
| | | Org Units | + | - | - |
| | | Interaction Protocols | - | - | - |
| | | Organizational structures | Plug-in "Social Network miner" | - | - |
| | Process Perspective | | Plug-ins "Alpha algorithm"," Genetic mining", "Multi-phase mining" | Induction and tranformation methods | Patterns Analyser |
| Capacity to support Multi-Analysis of processes | | Delta | Plug-in "Trace comparaison" | - | - |
| | | Performance | Plug-in "a Basic statistical analysis | - | Performance Analyser |
| Particularities of system | | Usability of interfaces | ++ | +/- | +/- |
| | | Portability | Java | - | Java |
| | | Extensibility | FrameWork | - | - |

### III. OUR PROPOSED WORKFLOW LOG META-MODEL

#### A. Motivation for using MAS approach

We think that multi-agent system approach can appropriately be accommodated for the supporting of the major drawback of existing WM systems and namely the discovering of organizational perspective. Indeed, the FIPA-ACL *performatives* (**F**oundations of **I**ntelligent **P**hysical **A**gents-**A**gent **C**ommunication **L**anguage, http://www.fipa.org) define clearly the semantic of messages and namely the agent's intentions (delegate, subcontract, negotiate…). *The organizational structures* such as hierarchy, federation, society and so on can model the behaviour of the actors group. I.e. they describe the macro-level dimension of the coordination among actors in terms of externally observable behaviour, independent of the internal features of each participating component. *The interaction protocols* could be used with benefits to allocate activities between actors. To conclude, it should be noted that the organizational structures and interaction protocols that we intend to discover them from Workflow log have been deeply investigated in Multi-Agent System (MAS) area and as a consequence it is possible to benefit from the results obtained in this area. To the best of our knowledge, the Multi-Agent approach has been widely used to study and implement workflow but has never been used in workflow mining. The interested reader can find more information about the feasibility of the agent approach in ([10] and [11]).

#### B. The proposed Workflow log meta-model

This meta-model is shown in the UML diagram of Figure 1.

In this UML Meta-model, a Process is composed of one (or several) Process Instance(s). Each Process Instance is composed of one (or several) Event Line(s). Each Event makes references to the following elements:

*An Activity* which is described through the Act_Name, EventStream and TimeStamp Attributes,

*A Document* which is described through the Doc_Name Attribute,

*An Actor* which is described through the Actor_ID and Actor_Name attributes. It's a member of organizational unit and plays one (or several) Role(s),

*A Role* which is described through the Role_Name Attribute,

*An Organizational Unit* which is described through the Org_Unit_Name Attribute,

*A Performative* which is described through the Per_Name Attribute.

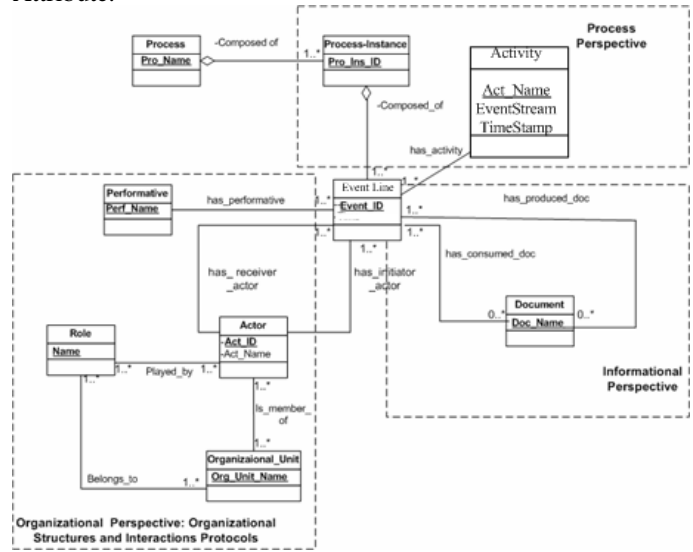

Figure 1. The proposed Workflow log Meta-model

Besides, the consumed documents and the produced documents are respectively represented by the Has_Consumed_Doc and Has_Produced_Doc relation Ships. The initiator Actor and the receiver actor are represented respectively by the Has_Initiator_Actor and Has_Receiver_Actor relations ship.

The process perspective is described with the Process Instance, EventLine and Activity classes. The informational perspective is described with the Process Instance, EventLine and Document classes. The organizational perspective is represented with Actor, Role, Organizational Unit and Performative classes.

To better illustrate our approach to deal with discovering organizational structures and interaction protocols, we give here an extract of the log file describing the management process of water distribution crisis (see table 2). For more information about this case study, the reader can refer to this work [11].

According to this table, we can discover easily the strict hierarchy structure and the contract net protocol.

TABLE II. An extract of the log file.

| Case | Performative | Activity | Initiator | Receiver |
|------|--------------|----------|-----------|----------|
| C1 | Execute | Investigation report establishment | Mahdi | System |
| C1 | Delegate | Alarm triggering | Mahdi | Salim |
| C1 | Inform | Alarm triggering | Salim | Mahdi |
| C1 | Cfp | Analysis of samples | Malik | Sami |
| C1 | Cfp | Analysis of samples | Malik | Amal |
| C1 | Propose | Analysis of samples | Amal | Malik |
| C1 | Propose | Analysis of samples | Sami | Malik |
| C1 | Accept-proposal | Analysis of samples | Malik | Amal |
| C1 | Reject-proposal | Analysis of samples | Malik | Sami |
| C1 | Execute | Analysis of samples | Amal | System |

More precisely, the second line of the log file describes the strict hierarchy structure between the actors Mahdi and Salim. At the same case C1, the grey part of this file shows the employed contract net protocol among the actors Malik, Amal and Sami. Indeed, the Malik actor proceeds by a Call for proposal (Cfp) to deal with the "Analysis of samples" activity. While the other actors Amal and Sami propose their bid. Finally, Malik notifies each participant either by acceptation (Accept-proposal) or by rejection (Reject-proposal).

## IV. THE DISCOPFLOW TOOL

This section is dedicated to the presentation of DiscopFlow tool for supporting organizational structures and interaction protocols discovering in workflow. More precisely, we first present its general architecture and then we expose the interface of DiscopFlow.

### A. Gneral arhitecture

The DiscopFlow tool, shown in figure 2, bases on the following architecture. More precisely, the DiscopFlow includes seven modules and one data base. The developed modules with eclipse platform are InterPro Analyser, OrgStruct Analyser, Info Analyser, AGR Analyser and Performance Analyser.

The Workflow log data base is developed with Oracle 10g according to the object relational model. Let us detail each module.

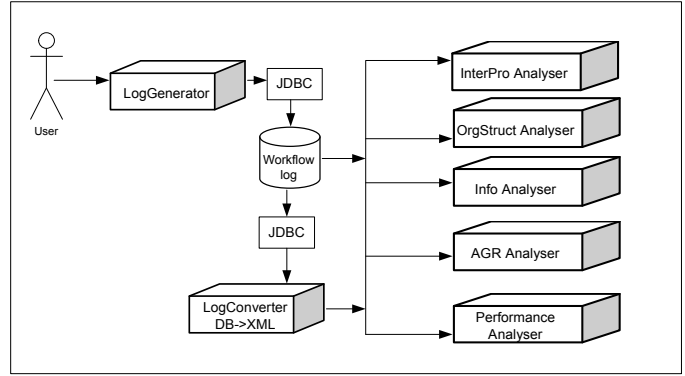

Figure 2. The general architecture of DiscopFlow tool

`LogGenerator`. It permits to generate automatically the workflow log according to the proposed workflow log meta-model under form data base.

`LogConverter`. It permits to convert a workflow log from database to XML schema. This module is in progress.

`OrgStruct Analyser`. It permits to discover the organizational structures or social networks such strict hierarchy, relaxed hierarchy, federation, coalition and so on,

`InterPro Analyser`. It supports the discovering of interaction protocols such as contract net, auction, vote and so on.

`Info Analyser`. It permits the discovering of consumed and produced documents.

`AGR Analyser`. It gives a graphic representation of each workflow actor according AGR model,

`Performance Analyser`. It gives a statistical data about the execution average time of activity, the number of suspended/achieved activities by a given actor, correlation between interaction protocol and event stream of an activity and so on. This module is also in progress.

The OrgStruct analyzer and the InterPro Analyser are based on algorithms that we have defined. The interested reader can find more information about these algorithms in [12]

### A. Implementation

This work has been implemented as a part of the DiscopFlow project, whose objective is to support the discovering of the three complementary Workflow perspectives such as the organizational perspective, the informational perspective and the process perspective.

The current version of DiscopFlow aims at discovering the organizational perspective and notably the discovering of organizational structures and interaction protocols.

DiscopFlow has been implemented using the Eclipse platform which allows the development of extensible applications using free plug-ins [13]. The figure 3 shows some screenshots of DiscopFlow.

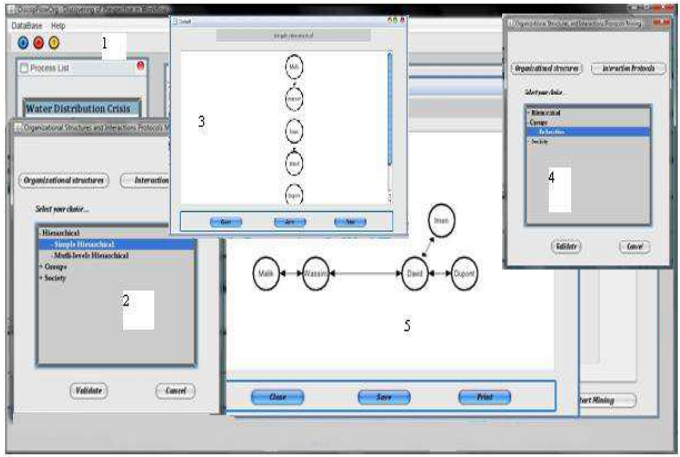

Figure 3. Some screenshots of DiscopFlow

The window number 1 presents the general menu of the DiscopFlow. This menu helps the user i) to connect to the data base implementing our enriched Workflow log meta-model and ii) have an idea about the main functionality of the DiscopFlow. This window also provides some icons buttons easing the discovering of process. Before starting the organizational perspective discovering, the user can see the general data about the Workflow log (business processes and instances), the organizational perspective (actors, performatives, organizational units and roles), the informational perspective (consumed documents and produced documents) and the process perspective (activities, timestamp and event stream). As mentioned previously, we only show the discovering of organizational structures and interaction protocols.

The windows 2 and 4 expose respectively the organizational structures (Strict Hierarchy, Relaxed Hierarchy, Federation, etc.) and interaction protocols (contract net, auction, argumentation, etc.) which can be discovered.

The windows 3 and 5 correspond respectively to the discovering of strict hierarchy and federation.

## V. DISCUSSIONS AND CONCLUSION

This paper has presented a critical and a comparative study of some existing WM systems according to quality criteria that we have defined. Even if these systems are powerful, they are not, in our opinion, completed since they neglect the important point that workflow is much more that process perspective. This paper has:

(i) defended the idea that the use of agent approach and in particular the organizational dimension in Multi-Agent System (MAS) is well suited to deal with discovering of organisational structures and interaction protocols in Workflow.

(ii) It specified a Workflow log meta-model having a multi-perspectives view such as informational perspective, process perspective and organizational perspective. This latter perspective integrates concepts inspired from agent approach and more precisely, it integrates Actors, roles, organizational units and FIPA-ACL performatives concepts.

(iii) it developed a tool called DiscopFlow which supports the discovering of organizational structures and interactions protocols.

Regarding related works, workflow mining has already given place to several works and systems such as ProM [4], InWoLvE [5] and WorkflowMiner [6]. The most of these works or systems mainly address workflow mining by considering only the process perspective.

For instance [4] proposes the ProM system which only supports the discovering of process and organizational perspectives. Regarding organizational perspective, it provides three methods such as default mining, mining based on the similarity of activities and mining based on the similarity of cases. On the one hand, this solution only supports the classical elements such as role and organizational unit and the social net work and notably the relaxed hierarchy. On the other hand, it does not exploit the agent approach and as a consequence it does not support the interaction protocols and organizational structures as defended in our work. [5] Proposes the InWolve system which creates in the first step a stochastic activity graph from workflow log and in the second step, it transforms the activity graph into a well defined process workflow model. [6] proposes the WorkflowMiner system which aims at discovering Workflow patterns from workflow log by using a statistical technique.

These works differ from our work for several reasons.

First, we provide a solution for discovering of organizational perspective and more precisely the organizational structures and interaction protocols mining. Second, we highlight how organizational dimension in multi-agent system can help the discovering of these organizational structures and interactions protocols by extending the classical workflow log. Besides, we provide an AGR (Agent, Group and Role) [14] graphic representation of each actor (i.e. its role and its organizational unit). Third, workflow mining is not limited to process perspective as we also consider organizational and informational perspectives.

As future work we plan to finish the development of Performances Analyser and LogConverter modules. We also plan to discover conjointly the three workflow perspectives.

## VI. APPONDIX

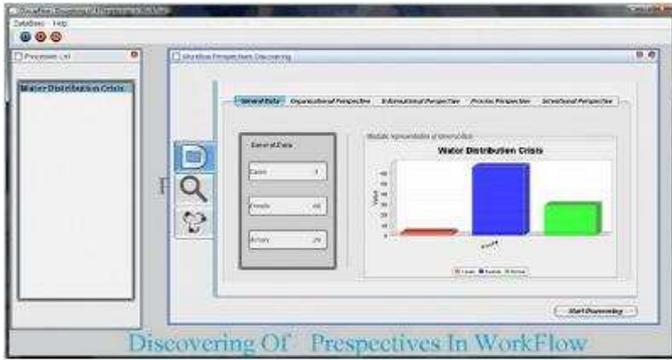

Figure 4. General date about the workflow log

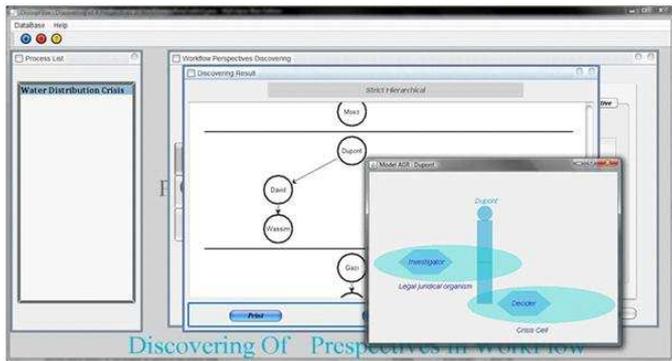

Figure 5. AGR representation of workflow actor

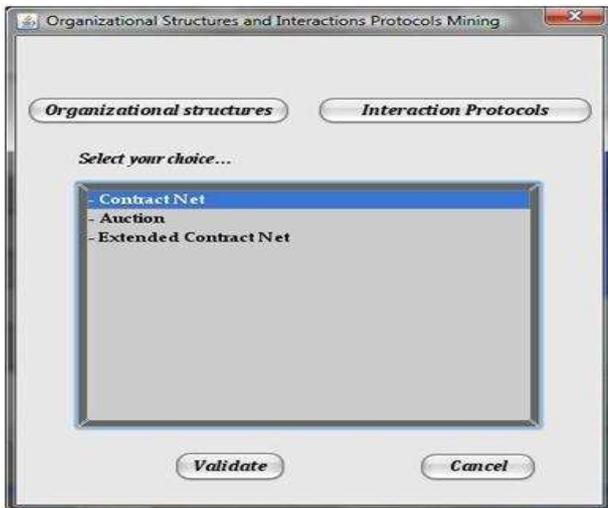

Figure 6. Choice of OS and IP to be discovered